**Title** pH sensing properties of flexible, bias-free graphene microelectrodes in complex fluids: from phosphate buffer solution to human serum

*Jinglei Ping[†], Jacquelyn E. Blum[‡], Ramya Vishnubhotla[†], Amey Vrudhula[§], Carl H. Naylor[†], Zhaoli Gao[†], Jeffery G. Saven[‡], & A. T. Charlie Johnson[\*,†]*

[†]Department of Physics and Astronomy, University of Pennsylvania, Philadelphia 19104, United States

Email: cjohnson@physics.upenn.edu

[‡]Department of Chemistry, University of Pennsylvania, Philadelphia 19104, United States

[§]Department of Bioengineering, University of Pennsylvania, Philadelphia 19104, United States

KEYWORDS: pH, graphene, flexible, microelectrode, tumor

ABSTRACT: Advances in techniques for monitoring pH in complex fluids could have significant impact on analytical and biomedical applications ranging from water quality assessment to *in vivo* diagnostics. We developed flexible graphene microelectrodes (GEs) for rapid (< 5 seconds), very low power (femtowatt) detection of the pH of complex biofluids. The method is based on real-time measurement of Faradaic charge transfer between the GE and a solution at zero electrical bias. For an idealized sample of phosphate buffer solution (PBS), the Faradaic current varied monotonically and systematically with the pH with resolution of ~0.2 pH unit. The current-pH dependence was well described by a hybrid analytical-computational model where the electric double layer derives from an intrinsic, pH-independent (positive) charge associated with the graphene-water interface and ionizable (negative) charged groups described by the Langmuir-Freundlich adsorption isotherm. We also tested the GEs in more complex bio-solutions. In the case of a ferritin solution, the relative Faradaic current, defined as the difference between the measured current response and a baseline response due to PBS, showed a strong signal associated with the disassembly of the ferritin and the release of ferric ions at pH ~ 2.0. For samples of human serum, the Faradaic current showed a reproducible rapid (<20s) response to pH. By combining the Faradaic current and real time current variation, the methodology is potentially suitable for use to detect tumor-induced changes in extracellular pH.

# 1. Introduction

*In vivo* monitoring of pH is important in investigations of tissue metabolism, neurophysiology, and diagnostics[1]. Extracellular pH-sensing, though of great interest for cancer diagnosis and medical treatment[1-4], is currently based mainly on relatively slow fluorescent techniques such as fluorogenic pH probes[5, 6] and fluorophore-decorated micelles[7]. Moreover, although optical methods hold promise for *in vivo* applications, improvement in detection platforms is still needed.[8] Other methods for *in vivo* measurement of tumor pH, including positron emission tomography (PET) radiotracers, magnetic resonance (MR) spectroscopy, and magnetic resonance imaging (MRI), are limited in sensitivity and require expensive instrumentation and exogenous and even radioactive indicators.[8] Electrical or electrochemical devices have the potential to be developed for *in vivo* pH monitoring but they are typically based on metal and glass, making them fragile and bulky. Existing approaches have additional disadvantages including the need for frequent recalibration, excessive power consumption, and lack of biocompatibility[1].

Flexible field-effect transistors (FETs) based on graphene, a biocompatible[9], chemically inert, and scalable[10] two-dimensional material with high quality pH-sensing properties[11-19], are promising for monitoring pH changes in biological systems. One important application is in cancer research and diagnostics since tumors demonstrate substantial reduction in extracellular pH[2, 20, 21] by 1.5 pH unit (from ~7.5 for healthy tissue to ~6.0 for tumor) but only moderate fluctuations in sodium concentration (~ 7%)[22] with respect to normal tissue. However, graphene FETs are commonly operated with ~ 100 mV source-drain bias and ~ 400 mV liquid-gate voltage. The application of these potentials/biases may complicate device fabrication, scaling, and stability; perturb the system under investigation; and set a power (and thus size) constraint on the device. Since each gate-sweep measurement requires ~100 seconds to identify the charge neutrality point that characterizes the pH value, the pH measurement process with a FET is also relatively slow and may not be suitable for real-time monitoring.

Here we demonstrate the use of flexible graphene microelectrodes (GEs) [23] for rapid, bias-free pH measurement in phosphate buffer solution (PBS), ferritin solution in PBS (0.1 μM), and human serum. The GE fabrication process is based on scalable photolithographic approaches, and the measurements are conducted without using an external bias voltage [23], so the methodology is intrinsically low-power and minimally perturbative. We find that the spontaneous Faradaic charge transfer between the GE and PBS is modulated by the pH. The Faradaic current extracted from 5 seconds of charge measurement (20 times faster than graphene FETs[11-19]) varies systematically with the pH of PBS and is very insensitive to moderate fluctuations of the extracellular ionic strength that would be induced by a tumor (~7%). The GE response to pH is well described by a hybrid analytical/computational model where the electric double layer derives from an intrinsic, pH-independent (positive) charge associated with the graphene-water interface and ionizable (negative) charged groups described by the Langmuir-Freundlich adsorption isotherm. For the ferritin solution, we focus on the relative Faradaic current obtained by subtracting the baseline Faradaic current for PBS from that for the ferritin solution. The relative Faradaic current shows a very strong feature that we associate with the disassembly of the ferritin cage and the associated release of ferric ions into the solution. For human serum, the GE reaches equilibrium with the solution in short time (~20 s) and also demonstrates remarkable performance: the Faradaic current responds systematically to pH in the range from 6.0 to 7.6 with high resolution (<0.2 pH unit); the differential current with respect to the pH flips sign and changes by ~ 150% as the pH decreases from 7.1 to 6.4. Together these findings suggest the suitability of the GE for both monitoring of biomolecular activity or protein disassembly in solution and for measurement of pH reduction expected for tumor extracellular fluid (1.5 pH unit)[2, 20, 21] *in vitro* or *in vivo*.

## 2. Results and Discussion

### 2.1. Device Fabrication and Setup

Inch-size graphene sheets for scalable electrode fabrication[24] were synthesized via low-pressure chemical vapor deposition on copper[25], and then transferred onto a flexible Kapton polyimide film with a pre-fabricated array of gold contacts. An $Al_2O_3$ sacrificial layer was deposited onto the sample by e-beam evaporation, and then the GE structures were defined using photolithography and oxygen plasma etching. This was followed by spin-coating of a 7-μm thick SU-8 2007 (Microchem) passivation layer, which was patterned to define 100 μm × 100 μm wells over the graphene electrodes. (See Experimental Section for further details of the device fabrication) An example of as-fabricated flexible devices is shown in **Figure 1a**.

The sub-pA Faradaic current between the GEs and the solution under test was measured using an electrometer (Keithley 6517a) with high resolution (~ fC) and low noise (0.75 fC/s peak-to-peak), as shown in **Figure 1b**. The noninverting input of the electrometer was initially grounded. The GE was exposed to fluid samples with various pH values. To conduct the measurement, the graphene electrode was connected to the inverting input of the operational amplifier of the electrometer, and the charge transferred from the solution to graphene accumulated on the feedback capacitor $C_f$ to provide the readout of the electrometer.

**2.2. Modulation of Faradaic Current through pH Variation**

First, we monitored the Faradaic charge transfer as a function of time for PBS (ionic strength 150 mM) as the pH was decreased from 11.2 to 2.2 and then increased back to 7.1 (**Figure 2a**). For each pH value, the charge transferred from the solution to the graphene increased linearly with time, with the slope used to determine the Faradaic current $i$. In contrast to gate-sweep measurements for graphene FETs, where several minutes might be needed to determine the shift in Dirac voltage that indicates the pH, the Faradaic current measurement described here was completed in less than 5 sec. The Faradaic current decreases monotonically with increasing pH, with excellent reproducibility, and minimal hysteresis (**Figure 2b**). For pH > 3, the Faradaic current is negative, i.e., electrons are transferred from the solution to the graphene. At low pH (<3), the Faradaic current is positive indicating that the proton concentration in the solution is large enough to reverse the direction of the current. The dependence of the Faradaic current on pH is approximately (but not exactly) linear, with a sensitivity of ~0.12 ± 0.01 pA/pH for pH in the range 2.2 – 11.2. We get an excellent fit to the data (red curve in **Figure 1b**) using a model that incorporates the electric double layer and ionizable defect groups on graphene, as described in the following paragraphs.

The equivalent circuit model[26] describing the graphene-solution interface is shown in the inset of **Figure 2b**. The interfacial capacitance $C_i$ (~ µF cm$^{-2}$)[27] of the graphene/solution interface can be ignored in the DC measurement used here, so the Faradaic current $i$ is determined by the electrostatic potential $\psi_S$ of the Stern plane due to adsorbed charges near the graphene surface and the charge transfer resistance $R_{ct}$ (~ 6.7 MΩ cm$^2$)[23] between the graphene and the ionic solution: $i = \psi_S/R_{ct}$. The measured sensitivity of the GE, 0.12 ± 0.01 pA/pH, is equivalent to 6.8 ± 0.7 mV/pH at the Stern plane, in good agreement with the value[16] of ~ 6 mV/pH reported by others for experiments on graphene FETs in ionic aqueous solution using an electrolytic gate.

The current-pH dependence can be fit quantitatively using a model where the Grahame equation[28] is used to quantify the potential at the Stern plane associated with a surface charge density, $\sigma_S$, with two components: a constant (i.e., pH-independent) offset charge density and a set of ionizable defect sites in the graphene whose charge state varies with proton concentration through the the Langmuir-Freundlich isotherm[13, 14]:

$$i = \psi_S/R_{ct} \quad (1)$$

$$\psi_S = \frac{2k_B T}{e} \sinh^{-1} \frac{\sigma_S}{\sqrt{8\epsilon\epsilon_0 k_B T c}} \quad (2)$$

$$\sigma_S = \frac{\sigma_{max}}{1+10^{n(pK_a-pH)}\exp(-ne\psi_S/k_B T)} + \sigma_{off} \quad (3)$$

In **Equation 2**, $k_B$ is the Boltzmann constant, $T$ the absolute temperature, $e$ the electronic charge, $\epsilon$ ($\epsilon_0$) the relative (vacuum) permittivity, and $c$ = 150 mM the ionic strength of the solution. In **Equation 3**, $\sigma_{max}$ is the areal charge density of ionizable groups (i.e., graphene defects), pK$_a$ is the dissociation constant, and $n$ the degree of heterogeneity.

The parameter σ$_{off}$ allows for the presence of a surface charge density that is independent of pH.

Combining **Equation 1-3**, we obtain an excellent fit to the measured current-pH response, where $\sigma_{max}$, $n$, p$K_a$, and σ$_{off}$ are the fit parameters (solid line in **Figure 2b**). The best fit value for $\sigma_{max}$ is - 0.077 ± 0.005 C m$^{-2}$, consistent with earlier reports for graphene and carbon nanotubes[19, 29, 30] with values in the range from - 0.01 to - 0.08 C m$^{-2}$. The best fit value for σ$_{off}$ is 0.007 ± 0.002 C m$^{-2}$, which we discuss further in the following paragraph. The best fit values for $n$ = 0.24 ± 0.03 and p$K_a$ = 6.5 ± 0.1 show reasonable agreement with values of $n$ = 0.3 and p$K_a$ = 7.6 found by others for single-wall carbon nanotubes (SWCNs) in KCl solution[30]. Our value for p$K_a$ is also in the range 4.3 – 9.8 from earlier reports for ionizable groups on graphene[31].

The pH-independent areal charge density σ$_{off}$ characterizes an intrinsic electric double layer at the graphene-water interface. To provide a molecular basis for this quantity, we first simulated the distribution of water molecules associated with defect-free graphene in contact with pure water with molecular dynamics. The charge density obtained from the simulation was then used to calculate the potential difference as a function of distance from the graphene (**Figure 2c**). (See Simulation Section for details.) At the graphene surface, a potential of $\Phi$ = +360 mV is calculated relative to bulk water. Excess hydrogen density compared to oxygen density close to the graphene surface (z < 0.3 nm) leads to the positive potential. Considering the double-layer capacitance at the graphene-solution interface, ~ 1.3 μF cm$^{-2}$ (assuming the hydrogen-graphene distance of 0.12 nm),[32] the corresponding charge density is approximately 0.005 C m$^{-2}$, in good agreement with the value of σ$_{off}$ inferred from the experiment (0.007 ± 0.002 C m$^{-2}$).

## 2.3. pH Response of Graphene Electrode to Complex Biofluids

Building on this understanding of GE operation in an idealized PBS sample, we conducted experiments to explore the use of GEs in more complex biological solutions. As a first step, we used a GE to measure the Faradaic current as a function of pH in a 0.1 µM equine spleen ferritin (Sigma Aldrich F4503) solution in PBS. Ferritin is a globular protein complex of 24 subunits found in most tissues and in serum (pH ~ 7.0) that stores iron oxide and releases it in a controlled fashion. Ferritin is known to disassemble and release the stored iron ions for pH below ~ 2.0 [33], with partial disassembly beginning to occur for pH below 4.2 [34]. Since kapton degrades at low pH below 2.0 [35], the GEs for this experiment were fabricated on oxidized silicon substrates. First we measured the Faradaic current for the ferritin solution as a function of pH over the range 1.0 – 7.0, and then we conducted the same measurement for a pure PBS solution to determine a baseline response (data not shown). The pH of all solutions was adjusted in steps of ~ 1.0 pH unit by adding 150 mM hydrochloride acid solution.

In order to observe the signature of ferritin disassembly, we focused on the *relative* Faradaic current (Figure 3), obtained by subtracting the baseline Faradaic current for PBS from that for the 5 µM ferritin solution. The relative Faradaic current increases abruptly at pH near 2.0, exactly in the range where ferritin disassembles and positively charged iron ions enclosed in the intact globular ferritin 24-mer are released. Furthermore, there is a noticeable increase in the relative Faradaic current in the pH range 2.0 - 4.0, in agreement with the expectation that partial disassembly of horse spleen ferritin occurs in this range.[34] Thus the pH dependence of the relative Faradaic current for the ferritin solution, although not analytically interpretable, is a sensitive probe of biomolecular processes that substantially change the electrostatic environment of the GE.

To test the GE performance in a complex human bio-fluid, we investigated its response to pH changes in a sample of human serum (ThermoFisher) diluted with PBS to bring it to physiological ionic strength ~ 150 mM. The pH range tested was 6.0 (typical extracellular pH for a tumor) to 7.6 (typical for normal tissue), which covers the range of pH variation that can be induced by non-metastatic/metastatic tumor.[2, 20, 21] The GE Faradaic current was measured over the same pH range in PBS at ionic strengths of 139.5mM, 150 mM, and 160.5 mM (**Figure 4a**), corresponding to the variation ionic strength expected in extracellular fluid (~7%)[22]. Over the relevant pH range, the Faradaic current varied by nearly 0.3 pA (~ 45%), with an estimated pH resolution < 0.2 pH unit and sensitivity of 0.144 ± 0.0098 pA/pH. For fixed pH, the variation of Faradaic current over the range of ionic strengths tested was only 0.01 – 0.02 pA, more than an order of magnitude smaller.

For human serum, the Faradaic charge transfer (**Figure 4b**) had a more gradual time dependence than that for PBS (**Figure 2a**). The variation of the GE Faradaic current with time in serum was well described by simple relaxation with a single time constant $\tau$ to a constant value that we term the steady-state Faradaic current (**Figure 4b**). At a pH of 7.60 (**Figure 4b,c**), the time constant was $\tau = 3.81 \pm 0.09$ s, and over the range of pH tested, this time constant varied by ± 0.5 s. This relaxation time is presumed to reflect equilibration processes such as non-specific adsorption of organic and inorganic components in human serum[36] onto the graphene surface, in rough agreement with earlier reports of the saturation time scale for non-specific adsorption of protein onto graphene (~ 30 s) measured with graphene FETs[12].

The magnitude of the Faradaic current measured in serum (Figure **4d**) is smaller by 0.1 – 0.4 pA over the whole pH range than that for PBS. The reduced current magnitude is ascribed to the inhibition of electronic communication to the GE by biomolecules adsorbed onto its surface [37, 38].

The differential current with respect to the pH ($\Delta I/\Delta[pH]$) can be derived from the current-pH response (Figure 4**a**), with results shown in Figure **4d**. The differential current response shows two different behaviors over the tested range: it is positive for pH 6.1 to 6.6 (saturating at ~ 0.47 pA/pH), and it is negative for pH 6.6 to 7.6 (saturating at ~ -0.23pA/pH) with an abrupt transition at pH ~ 6.6. Since tumor development almost exclusively leads to acidosis with very rare exceptions,[39] tumor-induced pH decrease will result in either Faradaic current reduction in the pH range of 6.1 to 6.6 or increase in the range of 6.6 to 7.6 (Figure 4**e**). Thus for tumor diagnosis, the range of the pH can be determined from the current variation and further the pH can be identified based on the current-pH response.

## 3. Conclusion

In summary, we have demonstrated the use of flexible graphene microelectrodes for monitoring of pH in idealized and complex bio-solutions, specifically PBS, 5 μM ferritin solution, and human serum. The measurement signal is the zero-bias, sub-pA Faradaic current between the GE and the solution, making this a low-power, minimally perturbative approach. For PBS, the variation of the current with pH can be understood quantitatively in a model where the current reflects the potential of the Stern layer, which is derived from an intrinsic (positive) charge associated with the graphene-water interface and ionizable (negative) charged groups whose density is described by a Langmuir-Freundlich adsorption isotherm. The charge density intrinsic to the graphene-water interface derived from the model is in excellent agreement with that found *via* molecular dynamics simulation. For the ferritin solution, the relative Faradaic current, compared to a PBS baseline, shows a strong feature at pH ~ 2.0, reflecting the disassembly of the ferritin cage and release of iron atoms. For human serum, the microelectrode rapidly (~ 20 s) reaches equilibrium with the solution. The Faradaic current and the current variation together can be used for identifying pH changes on the scale of that induced by a tumor. This electrode-based technique is therefore potentially suitable for use as a miniature portable or implantable pH-sensor for early-stage cancer diagnosis.

**4. Experimental Section**

**Graphene growth** Copper foil (99.8% purity) was loaded into a four-inch quartz tube furnace and annealed for 30 minutes at 1050 °C in ultra-high-purity (99.999%) hydrogen atmosphere (flow rate 80 sccm; pressure of 850 mT at the tube outlet) for removal of oxide residues. Graphene was then deposited by low-pressure chemical vapor deposition (CVD) using methane as a precursor (flow rate 45 sccm, growth time of 60 min).

**Graphene device fabrication** Contacts (5 nm/40 nm Cr/Au) were pre-fabricated by photolithography on the device substrate (either a Kapton film or an oxidized silicon wafer). The graphene/copper growth substrate was coated with a 500 nm layer of

poly(methyl methacrylate) (PMMA; MicroChem). The PMMA/graphene/copper trilayer was immersed in a 0.1 M NaOH solution and connected to the cathode of a power supply so that the PMMA/graphene bilayer was peeled off the copper substrate by hydrogen bubbles generated between the bilayer and the copper. After thoroughly rinsing with DI-water, the PMMA/graphene bilayer was transferred onto the contact array on the device substrate. After the removal of PMMA with acetone, the film was annealed on a hot plate at 200 °C for 1 hour under ambient conditions.

A 5 nm $Al_2O_3$ sacrificial layer[40] was deposited onto the graphene by e-beam evaporation. Photolithography (AZ 5214 E, Microchem) was then used to define 100 μm x 100 μm graphene electrodes; the AZ MIF developer also removes the $Al_2O_3$ sacrificial layer, so unwanted graphene material could then be etched with an oxygen plasma. Next a passivation layer of 7-$\mu$m thick photoresist (SU-8; MicroChem) was spin-coated onto the sample, and windows were defined at the locations of the graphene electrodes. Finally, the $Al_2O_3$ sacrificial layer graphene electrodes was removed with AZ 422 MIF and the whole as-fabricated film was hard-baked at 200 °C on a hotplate for 2 hours.

**Biosample preparation** Equine spleen ferritin (Sigma Aldrich F4503) samples were prepared at 0.1 μM concentration in full PBS solution (ionic strength 150 mM). Delipidated and dialyzed human serum (ThermoFisher 31876) was diluted by 1.73 times in DI-water, resulting in ionic strength of ~ 150 mM. The pH for solutions of ferritin or human serum was adjusted by adding diluted chloride acid or sodium hydroxide solution.

**5. Simulation Section**

The simulations consisted of two sheets of graphene, generated by the Nanotube Builder plug-in of the Visual Molecular Dynamics software (VMD)[41], separated by 20 Å each in contact with atomistic water molecules. Periodic boundary conditions were used and the graphene sheets were positioned parallel to the *x-y* plane. The *x-y* dimensions of a periodic rectangular box were selected such that each sheet and its images formed a defect-free, continuous sheet of graphene. Each sheet had dimensions of 50.348 Å by 45.376 Å. The pair of parallel sheets was centered within the box in the z-dimension, and the distance between the two sheets was 20.000 Å. Each sheet contained 924 carbon atoms, the positions of which were constrained throughout the simulation. Water molecules were added with VMD's Solvate plug-in. Water was present above and below the sheets with a vacuum between them. The initial dimensions of the box were 51.577 Å by 46.794 Å by 119.942 Å and the system contained a total of 7153 water molecules.

The simulations were carried out in the NPT ensemble at 300 K and 1.0 atm. The area of the periodic box in the *x-y* plane was held constant, and the cell length in the *z* direction was allowed to vary. The CHARMM36[42-44] force field parameters were used with the NAMD software package[45]. The water model was the three-site TIPS3P model[46-48]. The charge on each hydrogen atom is + 0.417*e* and - 0.834*e* on each oxygen atom, where *e* is the elementary charge. Bond distances in water molecules were constrained using the SHAKE algorithm[49]. Temperature was controlled with a Langevin thermostat with a damping coefficient of 1.0 ps$^{-1}$. Pressure was controlled with a Langevin piston barostat[50,51] with a period of 200 fs and a damping time of 100 fs. The particle mesh Ewald method was used to calculate long-range electrostatics beyond 14.0 Å, with a grid spacing of 1.0 Å. A 2 fs time step was used. The system was minimized for 20,000 steps, then heated incrementally to 300 K in steps of 5 K and 50 K over 160 ps. The system was equilibrated for 200 ps, then run for 10 ns, with configurations sampled every 1 ps. For each configuration, the charge density of the water molecules was calculated as a function of distance from the plane containing the carbon atoms of graphene.

The electric potential ($\phi$) as a function of distance from graphene ($z$) was calculated from the charge density ($\rho$) using the Poisson equation[52, 53]:

$$\frac{d^2\phi(z)}{dz^2} = -\frac{\rho(z)}{\varepsilon_0}$$

where $\varepsilon_0$ is the vacuum permittivity, $8.854 \times 10^{-12}$ C m$^{-1}$ V$^{-1}$ (F/m) or $5.526 \times 10^{-5}$ $e$ nm$^{-1}$ mV$^{-1}$. This equation is integrated twice under the boundary conditions that electric field and potential are zero in bulk, to give

$$\phi(z) - \phi(z_0) = -\frac{1}{\varepsilon_0} \int_{z_0}^{z} dz' \int_{z_0}^{s} \rho(z'') \, dz''$$

$$= -\frac{1}{\varepsilon_0} \int_{z_0}^{z} (z - z') \rho(z') dz'$$

where the final expression is obtained using integration by parts. The bulk, reference value of $z_0$ used was 4.4 nm.

**Acknowledgement**

This work was supported by the Defense Advanced Research Projects Agency (DARPA) and the U. S. Army Research Office under grant number W911NF1010093. J.E.B. and J.G.S. acknowledge partial support from NSF CHE 1412496 and from the Penn

Laboratory for Research on the Structure of Matter (NSF DMR 1120901). This work used the Extreme Science and Engineering Discovery Environment (XSEDE), which is supported by NSF grant ACI-1053575, under grant TG-CHE 110041. R.V. and C.H.N. acknowledge support from NSF EFRI 2-DARE 1542879. A.V. is grateful for support through Penn's Rachleff Scholars Program. Z.G. acknowledges support through grant 1P30 ES013508 from the National Institute of Environmental Health Sciences, NIH. The contents of this paper are solely the responsibility of the authors and do not necessarily represent the official views of NIEHS, NIH.

Author Contributions

A.T.C.J. directed the research. J.P. proposed and designed the experiment, fabricated the graphene microelectrodes, and carried out the graphene microelectrode measurements. J.E.B. conducted the all-atom molecular dynamics simulations under the supervision of J.G.S. R.V and A.V prepared the graphene samples, and R.V., A.V., C.H.N., and Z.G. assisted with device fabrication. J.P., J.E.B., J.G.S., and A.T.C.J. wrote the manuscript, with input and approval from all the authors.

Reference

1. Korostynska, O.; Arshak, K.; Gill, E.; Arshak, A. *IEEE Sensors J.* **2008,** 8, (1), 20-8.
2. Vinnakota, K.; Kemp, M. L.; Kushmerick, M. J. *Biophys J.* **2006,** 91, (4), 1264-87.
3. Bizzarri, R.; Arcangeli, C.; Arosio, D.; Ricci, F.; Faraci, P.; Cardarelli, F.; Beltram, F. *Biophys J.* **2006,** 90, (9), 3300-14.
4. Tannock, I. F.; Rotin, D. *Cancer Res.* **1989,** 49, (16), 4373-84.
5. Tung, C.-H.; Qi, J.; Hu, L.; Han, M. S.; Kim, Y. *Theranostics* **2015,** 5, (10), 1166-1174.
6. Anderson, M.; Moshinikova, A.; Engelman, D. M.; Reshetnyak, Y. K.; Andreev, O. A. *Proc. Natl. Acad. Sci. U. S. A.* **2016,** 113, (29), 8177-8181.
7. Ma, X.; Wang, Y.; Zhao, T.; Li, Y.; Su, L.-C.; Wang, Z.; Huang, G.; Sumer, B. D.; Gao, J. *J. Am. Chem. Soc.* **2014,** 136, (31), 11085-11092.
8. Zhang, X.; Liu, Y.; Gillies, R. J. *J. Nucl. Med.* **2010,** 51, 1167-1170.
9. Chung, C.; Kim, Y.-K.; Shin, D.; Ryoo, S.-R.; Hong, B.; Min, D.-H. *Acc. Chem. Res.* **2013,** 46, (10), 2211-2224.

10. Lerner, M. B.; Matsunaga, F.; Han, G. H.; Hong, S. J.; Xi, J.; Crook, A.; Perez-Aguilar, J. M.; Park, Y. W.; Saven, J. G.; Liu, R.; Johnson, A. T. C. *Nano Lett.* **2014,** 14, (5), 2709-2714.
11. Sohn, I.-Y.; Kim, D.-J.; Jung, J.-H.; Yoon, O. J.; Thanh, T. N.; Quang, T. T.; Lee, N.-E. *Biosensors and Bioelectronics* **2013,** 45, (15), 70-6.
12. Ohno, Y.; Maehashi, K.; Yamashiro, Y.; Matsumoto, K. *Nano Lett.* **2009,** 9, (9), 3318-3322.
13. Mailly-Giacchetti; Hsu, A.; Wang, H.; Vinciguerra, V.; Pappalardo, F.; Occhipinti, L.; Guidetti, E.; Coffa, S.; Kong, J.; Palacios, T. *J. Appl. Phys.* **2013,** 114, (8), 084505.
14. Ang, P. K.; Chen, W.; Wee, A. T. S.; Loh, K. P. *J. Am. Chem. Soc.* **2008,** 130, (44), 14392-3.
15. Lee, M. H.; Kim, B. J.; Lee, K. H.; SHIN, I.-S.; Huh, W.; Cho, J. H.; Kang, M. S. *Nanoscale* **2015,** 7, (17), 7540-7544.
16. Fu, W.; Nef, C.; Knopfmacher, O.; Tarasov, A.; Weiss, M.; Calame, M.; Schonenberger, C. *Nano Lett.* **2011,** 11, (9), 3597-3600.
17. Cheng, Z.; Li, Q.; Li, Z.; Zhou, Q.; Fang, Y. *Nano Lett.* **2010,** 10, (102), 1864-1868.
18. Ristein, J.; Zhang, W.; Speck, F.; Ostler, M.; Ley, L.; Seyller, T. *J. Phys. D: Appl. Phys.* **2010,** 43, (34), 345303.
19. Heller, I.; Chatoor, S.; Männik, J.; Zevenbergen, M.; Dekker, C.; Lemay, S. *J. Am. Chem. Soc.* **2010,** 132, (48), 17149-17156.
20. Estrella, V.; Chen, T.; Lloyd, M.; Wojtkowiak, J.; Cornnell, H. H.; Ibrahim-Hashim, A.; Bailey, K.; Balagurunathan, Y.; Rothberg, J. M.; Sloane, B. F.; Johnson, J.; Gatenby, R. A.; Gillies, R. J. *Cancer Res.* **2012,** 73, (5), 1524-1538.
21. Webb, B. A.; Chimenti, M.; Jacobson, M. P.; Barber, D. L. *Nature Reviews* **2011,** 11, (9), 671-677.
22. Madelin, G.; Kline, R.; Walvick, R.; Regatte, R. R. *Scientific Reports* **2014,** 4, (4763), 1-7.
23. Ping, J.; Johnson, A. T. C. *Appl. Phys. Lett.* **2016,** 109, (1), 013103.
24. Lerner, M.; Matsunaga, F.; Han, G.; Hong, S.; Xi, J.; Crook, A.; Perez-Aguilar, J.; Park, Y.; Saven, J.; Liu, R. *Nano Lett.* **2014**.
25. Li, X.; Cai, W.; An, J.; Kim, S.; Nah, J.; Yang, D.; Piner, R.; Velamakanni, A.; Jung, I.; Tutuc, E.; Banerjee, S. K.; Colombo, L.; Ruoff, R. S. *Science* **2009,** 324, 1312.
26. Kuzum, D.; Takano, H.; Shim, E.; Reed, J. C.; Juul, H.; Richardson, A. G.; de Vries, J.; Bink, H.; Dichter, M. A.; Lucas, T. H.; Coulter, D. A.; Cubukcu, E.; Brain, L. *Nature Communications* **2014,** 5, (5259), 1-10.
27. Ping, J.; Xi, J.; Saven, J. G.; Liu, R.; Johnson, A. T. C. *Biosensors and Bioelectronics* **2015**.
28. Israelachvili, J., *Intermolecular and surface forces: Revised third edition*. Acadmeic Press: 2011.
29. Zuccaro, L.; Krieg, J.; Desideri, A.; Kern, K.; Balasubramanian, K. *Scientific Reports* **2015,** 5, (11794), 1-13.
30. Back, J. H.; Shim, M. *The Journal of Physicsl Chemistry B* **2006,** 110, (47), 23736-23741.
31. Yoon, J.-C.; Thiyagarajan, P.; Ahn, H.-J.; Jang, J.-H. *RSC Advances* **2015,** 5, (77), 62772-62777.
32. Ping, J.; Xi, J.; Saven, J. G.; Liu, R.; Johnson, A. T. C. *Biosensors and Bioelectronics* **2016,** in press.
33. Crichton, R. R.; Bryce, C. F. *Biochem. J.* **1973,** 133, (2), 289-299.
34. Kim, M.; Rho, Y.; Jin, K. S.; Ahn, B.; Jung, S.; Kim, H.; Ree, M. *Biomacromolecules* **2011,** 12, (5), 1629-1640.
35. DeIasi, R.; Russell, J. *J. Appl. Polym. Sci.* **1971,** 15, 2965-2974.
36. Krebs, H. A. *Biochemistry* **1950,** 19, 409-430.
37. Wang, C. H.; Yang, C.; Song, Y. Y.; Gao, W.; Xia, X. H. *Advanced Funcional Materials* **2005,** 15, (8), 1267-1275.
38. Kim, B. S.; Hayes, R. A.; Ralston, J. *Carbon* **1995,** 33, (1), 25-34.
39. Chong, W. H.; Molinolo, A. A.; Chen, C. C.; Collins, M. T. *Endocrine-related Cancer* **2011,** 18, R53-R77.
40. Hsu, A.; Wang, H.; Kim, K. K.; Kong, J.; Palacios, T. *IEEE Electron Device Lett.* **2011,** 32, (8), 1008-1010.
41. Humphrey, W.; Dalke, A.; Schulten, K. *Journal of Molecular Graphics* **1996,** 14, 33-38.
42. Best, R. B.; Zhu, X.; Shim, J.; Lopes, P. E.; Mittal, J.; Feig, M.; D., M. J. A. *J. Chem. Theory Comput.* **2012,** 8, 3257-3273.
43. MacKerellJr., A. D.; Feig, M.; Brooks, C. L. *J. Am. Chem. Soc.* **2004,** 126, 698-699.
44. MacKerell, J., A. D.; Bashford, D.; Bellott, M.; Dunbrack Jr., R. L.; Evanseck, J. D.; Field, M. J.; Fischer, S.; Gao, J.; Guo, H.; Ha, S.; Joseph-McCarthy, D.; Kunchnir, L.; Kuczera, K.; Lau, F. T. K.; Mattos, C.; Michnick, S.; Ngo, T.; Nguyen, D. T.; Prodhom, B.; Reiher, I., W. E.; Roux, B.; Schlenkrich, M.; Smith, J. C.; Stote, R.; Staub, J.; Watanabe, M.; Wiorkiewicz-Kuczera, J.; Yin, D.; Karplus, M. *J. Phys. Chem. B* **1998,** 102, 3586-3616.
45. Phillips, J. C.; Braun, R.; Wang, W.; Gumbart, J.; Tajkhorshid, E.; Villa, E.; Chipot, C.; Skeel, R. D.; Kale, L.; Schulten, K. *Journal of Computational Chemsitry* **2005,** 26, 1781-1802.
46. Brooks, B. R.; Bruccoleir, R. E.; Olafson, B. D.; Sttes, D. J.; Swaminathan, S.; Karplus, M. *Journal of Computational Chemsitry* **1983,** 4, (2), 187-217.
47. Jorgensen, W. L.; Chandrasekhar, J.; Madura, J. D.; Impey, R. W.; Klein, M. L. *J. Chem. Phys.* **1983,** 79, 926-935.
48. Neria, E.; Fischer, S.; Karplus, M. *J. Chem. Phys.* **1996,** 105, (5), 1902-1921.
49. Ryckaert, J. P.; Ciccotti, G.; Berendsen, H. J. C. *J. Comput. Phys.* **1977,** 23, 317-341.

50. Martyna, G. J.; Tobia, D. J.; Klein, M. L. *J. Chem. Phys.* **1994,** 101, (5), 4177-4189.
51. Feller, S. E.; Zhang, Y.; Pastor, R. W.; Brooks, B. R. *J. Chem. Phys.* **1995,** 103, (11), 4613-4621.
52. Gurtovenko, A. A.; Vattulainen, I. *J. Chem. Phys.* **2009,** 130, (21), 215107.
53. Tieleman, D. P.; Berendsen, H. J. C. *J. Chem. Phys.* **1996,** 105, (11), 4871-4880.

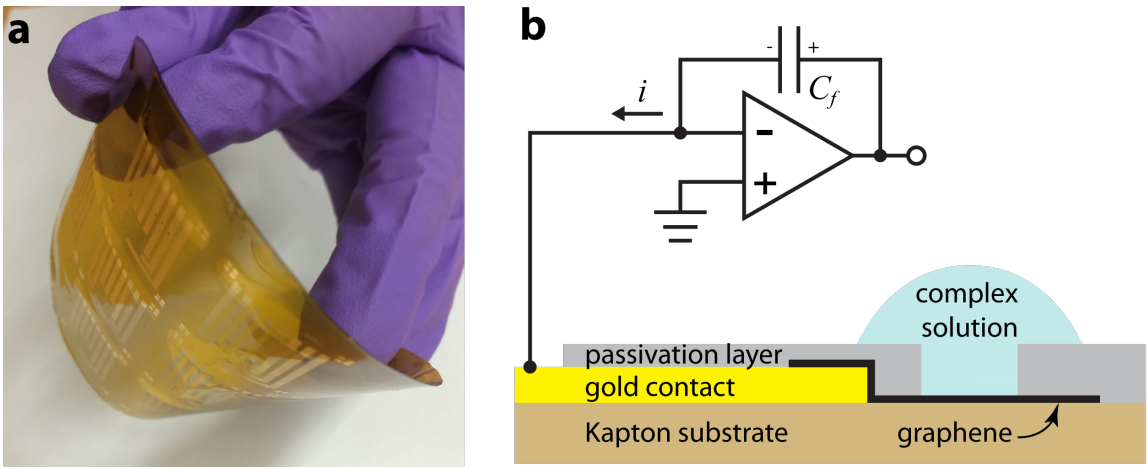

**Figure 1. a** Graphene electrode devices on a flexible polyimide substrate. **b** Schematic of the device and the measurement configuration.

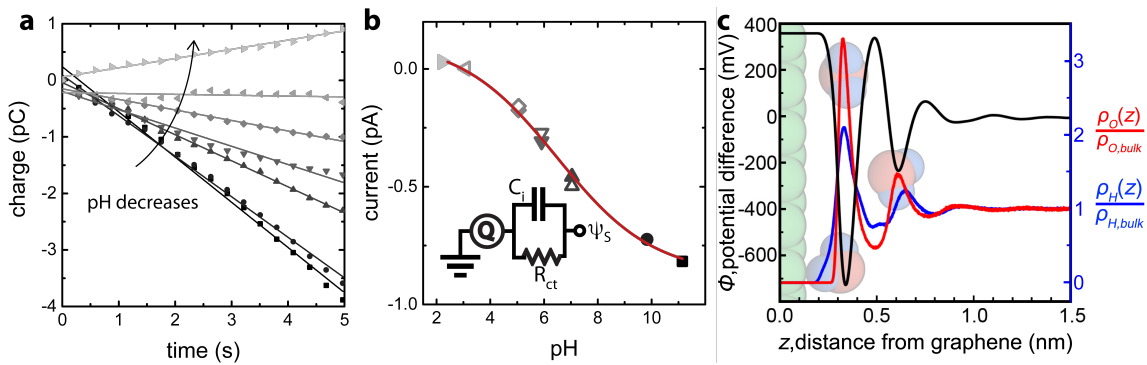

**Figure 2. a** Real-time Faradaic charge transfer for phosphate buffer solution of various pH values. The solid lines are linear fits to the data. **b** The Faradaic current extracted from **a** as a function of the pH. The starting pH was 11.2. The Faradaic current was measured as the pH was decreased to 2.2 (solid symbols) and then increased to 7.1 (open symbols). The solid curve is a fit to an equation derived from **Equations 1**-**3** in the main text. Inset: Equivalent circuit for the graphene-solution interface. **c** Molecular simulations were used to calculate electrostatic potential $\Phi(z)$ (black) and densities of water hydrogen atoms (blue) and water oxygen atoms (red) as functions of $z$, the distance from the plane containing the graphene carbon nuclei. The densities of oxygen $\rho_O$ and hydrogen $\rho_H$ are presented relative to the bulk values for these quantities, $\rho_{O,bulk}$ and $\rho_{H,bulk}$. Superimposed on the figure are space-filling representations of graphene (green) and representative configurations of water molecules at three different orientations and distances relative to the graphene surface; graphene and water molecules are rendered on the scale of the abscissa ($z$).

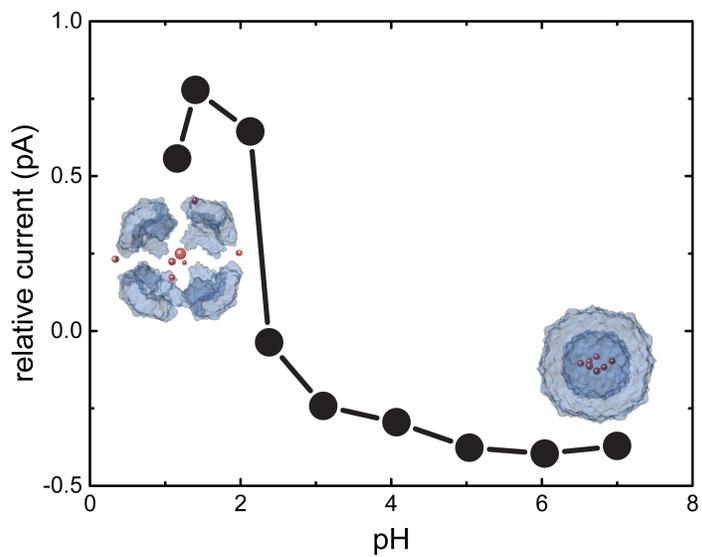

**Figure 3.** pH-dependence of the relative current for the ferritin solution, defined as the difference between the Faradaic current for the ferritin solution and the baseline current for PBS.

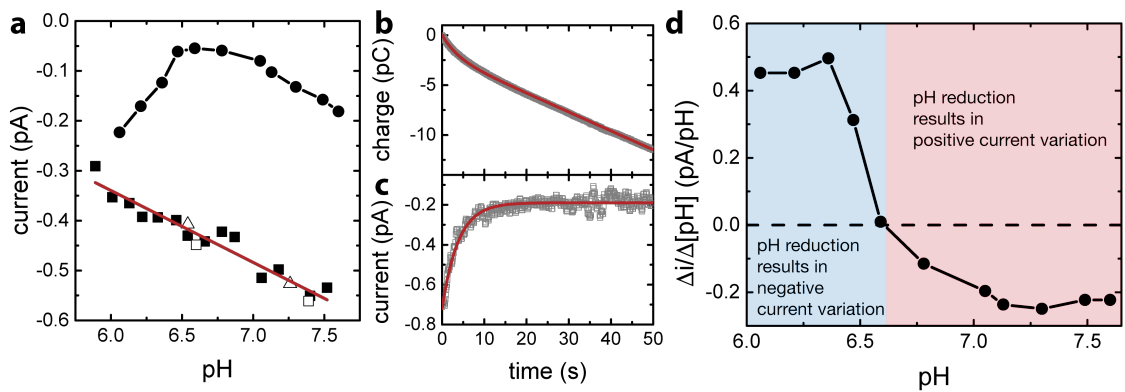

**Figure 4. a** Faradaic current for human serum sample diluted to ionic strength of 150.0 mM (solid circles) and for phosphate buffer solution (PBS) as a function of pH in the physiological range. PBS measurements were made at ionic strength values of 139.5 mM (hollow squares), 150.0 mM (solid squares), and 160.5 mM (hollow triangles). The red curve is a linear fit to the PBS data. **b**, **c** Time-dependence of the Faradaic charge transfer (panel **b**) and Faradaic current (panel **c**) for human serum at pH = 7.60. The red curves in **b** and **c** are fits to a model where the Faradaic current is described by a single relaxation time. **d** Differential current with respect to pH calculated based on the current response to serum in **a**.